\documentclass[letterpaper,11pt,twoside]{article}
\usepackage{amsmath,amssymb,amsthm, verbatim}
\usepackage{epic,eepic}

\newcommand{\xd}{\mathrm{d}}
\newcommand{\tens}{\otimes}
\newcommand{\defeq}{:=}

\newcommand{\C}{\mathbb{C}}

\DeclareMathOperator{\cou}{\epsilon}
\DeclareMathOperator{\cop}{\Delta}
\newcommand{\one}{\mathbf{1}}
\DeclareMathOperator{\id}{id}
\newtheorem{prop}{Proposition}
\newtheorem{thm}[prop]{Theorem}
\newtheorem{cor}[prop]{Corollary}
\newtheorem{lem}[prop]{Lemma}

\newcommand{\Rop}{R}
\newcommand{\Qop}{Q}

\newcommand{\Gcompl}{G}
\newcommand{\Gfeyn}{G_F}
\newcommand{\Gconn}{G_c}
\newcommand{\Gopi}{G_{\text{1PI}}}
\newcommand{\Gopim}{\hat{G}_{\text{1PI}}}

\newcommand{\compl}{\rho}
\newcommand{\conn}{\sigma}

\newcommand{\opi}{\tau}
\newcommand{\opim}{\hat{\tau}}
\newcommand{\SV}{\mathsf{S}(V)}

\newcommand{\Dopnew}{\Omega}
\newcommand{\lp}{l}
\newcommand{\vertex}{v}
\newcommand{\Top}{T}
\newcommand{\fct}{\nu}
\newcommand{\edge}{e}

\allowdisplaybreaks

\title{\textbf{Hopf algebras and the combinatorics of connected graphs in quantum field theory}}
\author{\^Angela Mestre\footnote{email: mestrang@kmlinux.fjfi.cvut.cz}\\
Doppler Institute \&
Department of Mathematics,\\
FNSPE,  Czech Technical University,\\
Trojanova 13, 120 00 Praha 2,\\
Czech Republic\\ \\
Robert Oeckl\footnote{email: robert@matmor.unam.mx}\\
Instituto de Matem\'aticas, UNAM, Campus Morelia,\\
C.~P.~58190, Morelia, Michoac\'an, Mexico\\
}

\begin{document}

\maketitle

\begin{abstract}
In this talk, we are concerned with the formulation and understanding of the combinatorics of time-ordered $n$-point functions in terms of the Hopf algebra of field operators. Mathematically, this problem can be formulated as one in combinatorics or graph theory. It consists in finding a recursive algorithm that generates all connected graphs in their Hopf algebraic representation. This representation can be used directly and efficiently in evaluating Feynman graphs
as contributions to the $n$-point functions.

\end{abstract}

Recently, it was realized that the Hopf algebra structure of the
algebra of field operators $\SV$ (with the normal or with the time-ordered
product) can be fruitfully exploited. In particular, the Laplace Hopf algebra created by  Rota {\em et al}. \cite{RSD:comb,RSG:ita} was generalized to
provide an algebraic tool for combinatorial problems of quantum field theory \cite{Br:qfa}. In addition,
 Sweedler's Hopf algebra cohomology \cite{Swe:cohom} and the Drinfel'd twist \cite{Dr}
 were used  to show that
different products of the algebra of field operators are related by
Drinfel'd twists and that interactions correspond to 2-cocycles \cite{BFFO:twist}. Subsequently,  the time-ordered Hoft algebra of  field operators  was  used to define an algebraic representation of graphs \cite{MeOe:npoint,MeOe:loop}. That is,  every graph with $\vertex$ vertices was associated with an element of $\SV^{\tens\vertex}$, the $\vertex$-fold tensor product of $\SV$.
This representation allowed to  derive simple algebraic relations between complete, connected and 1-particle irreducible (1PI) $n$-point functions \cite{MeOe:npoint}, and to express a connected $n$-point function in terms of its loop order contributions \cite{MeOe:loop}. The basic structure is an algorithm to recursively generate all weighted connected graphs. That is, each graph is generated together with a scalar that will correspond to the inverse of its symmetry factor.  This algorithm is amenable to direct
implementation and  allows efficient calculations of connected graphs as well as their values as Feynman graphs.   In particular, all results apply both to bosonic
and fermionic fields. Moreover, this Hopf algebraic approach to express the connected Green functions in terms of the 1PI ones, was recently generalized
to many-body physics \cite{BFP:1pi}.

This paper  reviews the main results of \cite{MeOe:npoint, MeOe:loop}.  Section \ref{sec:field} recalls the Hopf algebra structure of the algebra of quantum field operators. % $\SV$.
Section \ref{sec:algrep} describes the Hopf algebraic  representation of  graphs
and gives some examples.
Section~\ref{sec:loop} sketches
the algorithmic construction of connected graphs as well as the interpretation in terms of Feynman graphs and $n$-point functions.
Section \ref{sec:conn1pi} focuses on expressing the relation between connected and 1PI $n$-point functions in a completely algebraic language.

\section{Field operator algebra as a Hopf algebra}\label{sec:field}
We briefly recall the  Hopf algebra structure of the time-ordered field operator algebra. A more extensive discussion adapted to the present context, can be
found in \cite{MeOe:npoint}.

\bigskip

Let $V$ denote the vector space of linear combinations of elementary field
operators $\phi(x)$, where $x$ denotes field operator labels.
Let $\SV=\bigoplus_{\vertex=0}^{\infty}V^\vertex$ denote  the free, unital, commutative algebra generated by all  time-ordered products of field operators. In particular, $\SV$  is also a Hopf algebra \cite{Eisenbud, Kas, Loday}. Therefore, $\SV$ is equipped with a linear map
$\cop:\SV\to \SV\tens \SV$, called coproduct, which is coassociative: $(\cop\tens\id)\circ\cop=(\id\tens\cop)\circ\cop$.  This is defined on $V$ by
$\cop(\one) = \one\tens\one$,
$\cop(\phi(x)) =  \phi(x)\tens \one + \one\tens\phi(x)$,
and extended to the all of $\SV$ due to the compatibility with the product. The coproduct may be interpreted as an operation to split a product of
field operators into two parts in all possible ways. Moreover, on $\SV$ the counit
$\cou:\SV\to\C$ is defined by
$\cou(\one)=1$ and $\cou(\phi(x_1)\cdots\phi(x_n))=0$ for $ n>0$. The characterizing property of the counit is the equality
$(\cou\tens\id)\circ\cop=\id=(\id\tens\cou)\circ\cop$. Finally, to meet the requirements of a Hopf algebra,  there is an  antipode map $S:\SV\to\SV$ defined on $\SV$, by $S(\phi(x_1)\dots\phi(x_n))=(-1)^n \phi(x_1)\dots\phi(x_n).$

\smallskip

Now, let $\Gcompl^{(n)}$, $\Gconn^{(n)}$, $\Gopi^{(n)}$ and $\Gopim^{(n)}$  denote complete, connected, 1PI and  modified 1PI\footnote{Recall that the connected $n$-point functions $\Gconn$ are expressible
in terms of $\Gopim$ as a sum over all tree graphs whose vertices have valence at least three, by associating a
connected propagator $\Gconn^{(2)}$ to every edge.} $n$-point functions, respectively. Moreover, let $\mathcal{V}$ denote vertex functions. The ensemble of
time-ordered
$n$-point functions of a given type, determine  maps
$\SV\to \C$:
\begin{align*}
\compl(\phi(x_1)\cdots\phi(x_n)) & \defeq \Gcompl^{(n)}(x_1,\dots,x_n)\,,\\
\conn(\phi(x_1)\cdots\phi(x_n)) & \defeq \Gconn^{(n)}(x_1,\dots,x_n)\,,\\
\opi(\phi(x_1)\cdots\phi(x_n)) & \defeq \Gopi^{(n)}(x_1,\dots,x_n)\,,\\
\opim(\phi(x_1)\cdots\phi(x_n)) & \defeq \Gopim^{(n)}(x_1,\dots,x_n)\,,\\
\nu(\phi(x_1)\cdots\phi(x_n)) & \defeq \mathcal{V}(x_1,\dots,x_n)\,.
\end{align*}
The assumption that
all 1-point functions vanish means that
$\compl(\phi(x))=\conn(\phi(x))=\opi(\phi(x))=\opim(\phi(x))=0$.
Moreover,
the  0-point functions read as $\compl(\one)=1$,
$\conn(\one)=\opi(\one)=\opim(\one)=0$. Besides, the 2-point function
$\opim(\phi(x)\phi(y))$ vanishes by construction.

\section{A Hopf algebraic representation of graphs}
\label{sec:algrep}
We study the correspondence between graphs and elements of $\SV^{\tens\vertex}$ given in \cite{MeOe:npoint,MeOe:loop}.

\bigskip

Let $G_F^{-1}$ denote the inverse Feynman propagator given by  $$\int \xd y\, \Gfeyn(x,y)\Gfeyn^{-1}(y,z)=\delta(x,z)\,.$$
We consider the following formal operators defined on $\SV^{\tens\vertex}$:
\begin{equation}\label{eq:Rij}
\Rop_{i,j} := \int
\xd x\,\xd y\,G_F^{-1}(x,y)\,(\one^{\tens i-1}\tens \phi(x)
\tens\one^{\tens j-i-1}
\tens\phi(y)\tens\one^{\tens v-j}) ,
\end{equation}
where the field operators $\phi(x)$ and $\phi(y)$ are
inserted at the
positions $i$ and $j$, respectively, with $i\not=j$; and

\newpage 

\begin{equation}\label{eq:Rii}
\Rop_{i,i} := \int
\xd x\,\xd y\,G_F^{-1}(x,y)\,(\one^{\tens i-1}\tens \phi(x)
\phi(y)\tens\one^{\tens v-i})\,.
\end{equation}

\begin{figure}
\begin{center}
\setlength{\unitlength}{0.00083333in}
\begingroup\makeatletter\ifx\SetFigFont\undefined%
\gdef\SetFigFont#1#2#3#4#5{%
  \reset@font\fontsize{#1}{#2pt}%
  \fontfamily{#3}\fontseries{#4}\fontshape{#5}%
  \selectfont}%
\fi\endgroup%
{\renewcommand{\dashlinestretch}{30}
\begin{picture}(5112,1925)(0,-10)
\put(3555,1491){\ellipse{330}{496}}
\path(1516,1649)(1756,1124)
\path(1426,704)(1741,1229)
\path(2296,1094)(1681,1184)
\put(2401,1004){\makebox(0,0)[lb]{{\SetFigFont{12}{14.4}{\rmdefault}{\mddefault}{\updefault}$x_2$}}}
\put(1171,509){\makebox(0,0)[lb]{{\SetFigFont{12}{14.4}{\rmdefault}{\mddefault}{\updefault}$x_3$}}}
\put(1276,1754){\makebox(0,0)[lb]{{\SetFigFont{12}{14.4}{\rmdefault}{\mddefault}{\updefault}$x_1$}}}
\texture{44555555 55aaaaaa aa555555 55aaaaaa aa555555 55aaaaaa aa555555 55aaaaaa 
	aa555555 55aaaaaa aa555555 55aaaaaa aa555555 55aaaaaa aa555555 55aaaaaa 
	aa555555 55aaaaaa aa555555 55aaaaaa aa555555 55aaaaaa aa555555 55aaaaaa 
	aa555555 55aaaaaa aa555555 55aaaaaa aa555555 55aaaaaa aa555555 55aaaaaa }
\path(4506,1213)(5076,1213)
\path(4506,1213)(5076,1213)
\put(1725,1176){\shade\ellipse{240}{240}}
\put(1725,1176){\ellipse{240}{240}}
\put(128,1146){\shade\ellipse{240}{240}}
\put(128,1146){\ellipse{240}{240}}
\put(4984,1219){\shade\ellipse{240}{240}}
\put(4984,1219){\ellipse{240}{240}}
\put(4523,1222){\shade\ellipse{240}{240}}
\put(4523,1222){\ellipse{240}{240}}
\put(3551,1231){\shade\ellipse{240}{240}}
\put(3551,1231){\ellipse{240}{240}}
\put(76,52){\makebox(0,0)[lb]{{\SetFigFont{12}{14.4}{\rmdefault}{\mddefault}{\updefault}$\one$}}}
\put(1119,60){\makebox(0,0)[lb]{{\SetFigFont{12}{14.4}{\rmdefault}{\mddefault}{\updefault}$\phi(x_1)\phi(x_2)\phi(x_3)$}}}
\put(3353,60){\makebox(0,0)[lb]{{\SetFigFont{12}{14.4}{\rmdefault}{\mddefault}{\updefault}$\Rop_{1,1}$}}}
\put(4544,58){\makebox(0,0)[lb]{{\SetFigFont{12}{14.4}{\rmdefault}{\mddefault}{\updefault}$\Rop_{1,2}$}}}
\put(4941,1154){\makebox(0,0)[lb]{{\SetFigFont{12}{14.4}{\rmdefault}{\mddefault}{\updefault}2}}}
\put(1681,1109){\makebox(0,0)[lb]{{\SetFigFont{12}{14.4}{\familydefault}{\mddefault}{\updefault}1}}}
\put(83,1078){\makebox(0,0)[lb]{{\SetFigFont{12}{14.4}{\familydefault}{\mddefault}{\updefault}1}}}
\put(4475,1161){\makebox(0,0)[lb]{{\SetFigFont{12}{14.4}{\familydefault}{\mddefault}{\updefault}1}}}
\put(3506,1163){\makebox(0,0)[lb]{{\SetFigFont{12}{14.4}{\familydefault}{\mddefault}{\updefault}1}}}
\end{picture}
}
\caption{Examples of the correspondence between vertices and  edges,   and
 elements of $\SV^{\tens \vertex}$.}
\end{center}
\label{fig:dictionary}
\end{figure}
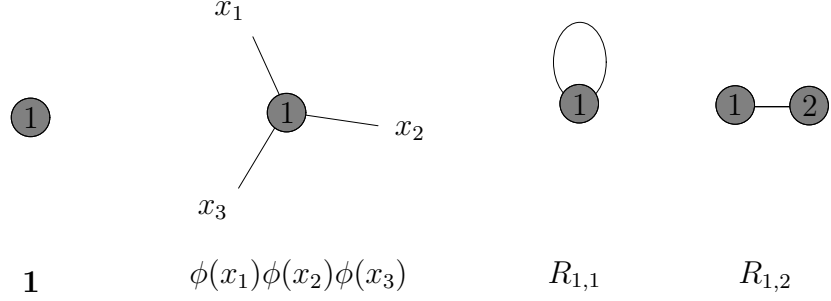

The $\Rop$-operators are employed in establishing a correspondence between graphs with $\vertex$
vertices and certain elements of $\SV^{\tens \vertex}$,
 the $v$-fold tensor product of $\SV$. Namely,
\begin{itemize}
\item
a  tensor factor in the $i^{\mbox{\tiny{th}}}$ position  corresponds to a
vertex
numbered  $i$, with $i=1,\dots,\vertex$;
\item
 a product
$\phi(x_1)\cdots\phi(x_n)$ in a
given tensor factor corresponds to external edges (of the associated
vertex) whose end points are labeled by $x_1,\dots,x_n$;
\item
the element $\Rop_{i,j}\in\SV^{\tens\vertex}$ defined by (\ref{eq:Rij}), corresponds to an
internal edge connecting
the vertices $i$ and $j$;
\item
the element $\Rop_{ i,i}\in\SV^{\tens\vertex}$ for $1\le i\le\vertex$ defined by (\ref{eq:Rii}),
corresponds to  an
internal edge connecting
the vertex $i$ to itself (i.e., a self-loop).
\end{itemize}
Figure  1 %\ref{fig:dictionary}
shows  examples of the correspondence between vertices, external edges and internal edges, and elements of $\SV^{\tens \vertex}$.

\begin{figure}
\begin{center}
\setlength{\unitlength}{0.00083333in}
\begingroup\makeatletter\ifx\SetFigFont\undefined%
\gdef\SetFigFont#1#2#3#4#5{%
  \reset@font\fontsize{#1}{#2pt}%
  \fontfamily{#3}\fontseries{#4}\fontshape{#5}%
  \selectfont}%
\fi\endgroup%
{\renewcommand{\dashlinestretch}{30}
\begin{picture}(6201,1999)(0,-10)
\put(4632,1285){\ellipse{1042}{234}}
\path(728,1251)(1658,1251)
\path(212,827)(782,1247)
\path(172,1682)(772,1247)
\path(1657,1295)(2227,1715)
\path(1634,1265)(2234,830)
\path(3527,1735)(4127,1300)
\path(3563,853)(4133,1273)
\path(5175,1288)(5745,1708)
\path(5201,1264)(5801,829)
\texture{44555555 55aaaaaa aa555555 55aaaaaa aa555555 55aaaaaa aa555555 55aaaaaa 
	aa555555 55aaaaaa aa555555 55aaaaaa aa555555 55aaaaaa aa555555 55aaaaaa 
	aa555555 55aaaaaa aa555555 55aaaaaa aa555555 55aaaaaa aa555555 55aaaaaa 
	aa555555 55aaaaaa aa555555 55aaaaaa aa555555 55aaaaaa aa555555 55aaaaaa }
\put(1628,1265){\shade\ellipse{240}{240}}
\put(1628,1265){\ellipse{240}{240}}
\put(766,1266){\shade\ellipse{240}{240}}
\put(766,1266){\ellipse{240}{240}}
\put(5161,1288){\shade\ellipse{240}{240}}
\put(5161,1288){\ellipse{240}{240}}
\put(4126,1296){\shade\ellipse{240}{240}}
\put(4126,1296){\ellipse{240}{240}}
\put(3443,634){\makebox(0,0)[lb]{{\SetFigFont{12}{14.4}{\rmdefault}{\mddefault}{\updefault}$x_2$}}}
\put(3841,80){\makebox(0,0)[lb]{{\SetFigFont{12}{14.4}{\rmdefault}{\mddefault}{\updefault}$\cdot(\phi(x_1)\phi(x_2)\otimes\phi(x_3)\phi(x_4))$}}}
\put(68,118){\makebox(0,0)[lb]{{\SetFigFont{12}{14.4}{\rmdefault}{\mddefault}{\updefault}$\Rop_{1,2}$}}}
\put(3518,58){\makebox(0,0)[lb]{{\SetFigFont{12}{14.4}{\rmdefault}{\mddefault}{\updefault}$\Rop^ 2_{1,2}$}}}
\put(435,110){\makebox(0,0)[lb]{{\SetFigFont{12}{14.4}{\rmdefault}{\mddefault}{\updefault}$\cdot(\phi(x_1)\phi(x_2)\otimes\phi(x_3)\phi(x_4))$}}}
\put(1585,1200){\makebox(0,0)[lb]{{\SetFigFont{12}{14.4}{\rmdefault}{\mddefault}{\updefault}2}}}
\put(15,1768){\makebox(0,0)[lb]{{\SetFigFont{12}{14.4}{\rmdefault}{\mddefault}{\updefault}$x_1$}}}
\put(0,675){\makebox(0,0)[lb]{{\SetFigFont{12}{14.4}{\rmdefault}{\mddefault}{\updefault}$x_2$}}}
\put(2049,1785){\makebox(0,0)[lb]{{\SetFigFont{12}{14.4}{\rmdefault}{\mddefault}{\updefault}$x_3$}}}
\put(2199,652){\makebox(0,0)[lb]{{\SetFigFont{12}{14.4}{\rmdefault}{\mddefault}{\updefault}$x_4$}}}
\put(5118,1223){\makebox(0,0)[lb]{{\SetFigFont{12}{14.4}{\rmdefault}{\mddefault}{\updefault}2}}}
\put(5740,612){\makebox(0,0)[lb]{{\SetFigFont{12}{14.4}{\rmdefault}{\mddefault}{\updefault}$x_4$}}}
\put(5643,1791){\makebox(0,0)[lb]{{\SetFigFont{12}{14.4}{\rmdefault}{\mddefault}{\updefault}$x_3$}}}
\put(3398,1828){\makebox(0,0)[lb]{{\SetFigFont{12}{14.4}{\rmdefault}{\mddefault}{\updefault}$x_1$}}}
\put(718,1205){\makebox(0,0)[lb]{{\SetFigFont{12}{14.4}{\familydefault}{\mddefault}{\updefault}1}}}
\put(4078,1235){\makebox(0,0)[lb]{{\SetFigFont{12}{14.4}{\familydefault}{\mddefault}{\updefault}1}}}
\end{picture}
}
\caption{Examples of the algebraic representation of graphs in terms
of elements of $\SV^{\tens \vertex}$.}
\end{center}
\label{fig:examples}
\end{figure}
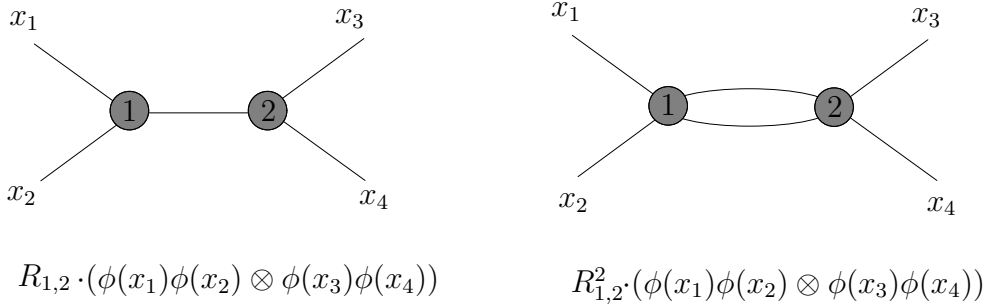

Combining several internal edges and their
products with external edges by multiplying the respective
expressions in $\SV^{\tens \vertex}$, allows to build arbitrary graphs
with $\vertex$ vertices. Figure \ref{fig:examples} shows some
examples.
Moreover, applying the vertex functions $\nu$
to each tensor factor yields precisely the
value of the respective graph as a Feynman graph.

Usually Feynman graphs involve
edges of different types depending on particle species, e.g. straight for fermions, wiggly for bosons, etc. Thus, suppose that there are  $m$ different fields $\phi^a$, with $a=1,\dots,m$,  interacting.  Each of these fields is associated with an edge of certain kind. In the present context, this means that   edges are represented  by distinct  elements $\Rop_{i,j}^a\,,\Rop_{i,i}^a\in \SV^{\tens\vertex}$, with $a=1,\dots,m$,  given by
$$\Rop_{i,j}^a := \int
\xd x\,\xd y\,{G_F^a}^{-1}(x,y)\,(\one^{\tens i-1}\tens \phi^a(x)
\tens\one^{\tens j-i-1}
\tens\phi^a(y)\tens\one^{\tens v-j})\,,$$
and
$$\Rop_{i,i}^a := \int
\xd x\,\xd y\,{G_F^a}^{-1}(x,y)\,(\one^{\tens i-1}\tens \phi^a(x)
\phi^a(y)\tens\one^{\tens v-i})\,,$$
respectively. Therefore,   the elements $\Rop_{i,j}$ and $\Rop_{i,i}$ that we consider read explicitly as
$\Rop_{i,j}=\sum_{a=1}^m\Rop^a_{i,j}$
and
$\Rop_{i,i}=\sum_{a=1}^m\Rop^a_{i,i}$,
respectively.

\smallskip

A fundamental property of the algebraic representation is that  the ordering of the tensor factors of $\SV^{\tens \vertex}$
induces an ordering of the vertices of the graphs.  However, when applying
$\fct^{\tens \vertex}$ the ordering is ``forgotten''. Indeed, it is not
relevant for the interpretation of graphs as Feynman graphs, but
only plays a role at
the level of their algebraic representation.
Moreover,  usually the elements of $\SV^{\tens \vertex}$
 yield as  linear combinations of expressions corresponding to
graphs. In this context, we call the scalar multiplying the expression
for a given graph the {\em weight} of the graph.
Clearly, if we are interested in unordered graphs, the  weight of such
a graph is the sum of the weights of all vertex ordered graphs that
correspond to it upon forgetting the vertex order.

\section{Generating connected graphs via Hopf algebra}
\label{sec:loop}
\subsubsection*{Statement of result}
We state the main result of \cite{MeOe:loop}. This may be described as an algorithm to recursively generate all connected graphs. In particular, each graph is produced together with a  weight factor given by the inverse of its symmetry factor. All graphs are generated in the Hopf algebraic representation introduced in Section \ref{sec:algrep}.  This allows their direct evaluation as Feynman graphs.

\smallskip

We notice that the discussion here applies to bare $n$-point functions only.

\bigskip

The elements
$\Rop_{i,j}$ and $\Rop_{i,i}$, given by formulas (\ref{eq:Rij}) and (\ref{eq:Rii}), respectively, are used to define the following linear maps:
\begin{itemize}
\item
$\Top_i:\SV^{\tens\vertex}\to\SV^{\tens\vertex}$,  with $1\le i\le\vertex$, as the operator $\Rop_{i,i}$ together with the factor $1/2$:
$$\Top_i\defeq\frac{1}{2}\Rop_{i,i}\,,$$
\item
$\Qop_i:\SV^{\tens\vertex}\to \SV^{\tens{\vertex+1}}$,  with $1\le i\le\vertex$, given by the composition of $\Rop_{i,i+1}$ with the coproduct applied to the $i^{\mbox{\tiny{th}}}$ component of $\SV^{\tens\vertex}$, i.e., $\cop_i\defeq
\id^{\tens{i-1}}\tens\cop\tens\id^{\tens{\vertex-i}}:
\SV^{\tens\vertex}\to\SV^{\tens{\vertex+1}}$,
together with the factor $1/2$:
$$\Qop_i \defeq \frac{1}{2} \Rop_{i,i+1}\circ\cop_i\,.$$
\end{itemize}
The map $\Top_i$  endows the
vertex $i$ of a vertex ordered graph with a self-loop.
The map $\Qop_i$ splits the vertex $i$ into two new vertices, numbered $i$ and $i+1$, distributes the ends of edges ending on the split vertex between the two new ones in all possible ways and connects the two new vertices with an edge.

The maps $\Top_i$ increase both the loop and edge
numbers of a graph by one unit, leaving the vertex number invariant. Also,
 the maps $\Qop_i$ increase both the vertex and edge numbers by
one unit, leaving the loop number invariant.

Clearly, both the maps $\Top_i$ and $\Qop_i$ produce connected graphs from connected ones, so that the following theorem holds.

\begin{thm}
\label{thm:main}
Let  $\lp, n\ge 0$, $\vertex\ge 1$ denote integers and let
 the set of maps
$\Dopnew^{\lp,\vertex}:\SV\to\SV^{\tens\vertex}$ be defined recursively as follows:
\begin{align}
\Dopnew^{0,1} & \defeq \id\,, \nonumber\\
\Dopnew^{\lp,\vertex} & \defeq
\frac{1}{\lp+\vertex-1}\left(\sum_{i=1}^{\vertex-1}\Qop_i \circ
\Dopnew^{\lp,\vertex-1}+\sum_{i=1}^{\vertex}
\Top_i\circ\Dopnew^{\lp-1,\vertex}\right)\,.
 \label{eq:recomega}
\end{align}
Then,  for fixed values of  $\lp,\vertex$, $n$ and  operator labels $x_1$,...,$x_n$, $\Dopnew^{\lp,\vertex}(\phi(x_1)\cdots\phi(x_n))$
corresponds to the weighted sum over all connected graphs with  $\lp$
loops, $\vertex$  vertices
and $n$ external edges whose end points are  labeled by
$x_1,\dots,x_n$,
each with weight the inverse of its symmetry factor.
\end{thm}

In the recursion equation above the $\Qop$ and $\Top$
summands do not appear when $\vertex=1$ or when $\lp=0$,
respectively.  The proof of Theorem \ref{thm:main} proceeds by induction on the number of internal edges $\edge=\lp+\vertex-1$ \cite{MeOe:loop}. Moreover, formula (\ref{eq:recomega}) is an example of a double recursion. Therefore, its algorithmic implementation  is that of any recurrence which makes two calls to itself, such as the  defining recurrence of the binomial coefficients. 

\smallskip

Now, we turn to the interpretation in terms of Feynman graphs
and $n$-point functions.
Denote the $\lp$-loop and $\vertex$-vertex
contribution to the ensemble $\conn$ of connected $n$-point functions
by $\conn^{\lp,\vertex}$. The $\lp$-loop order
contribution $\conn^\lp$ to $\conn$ and $\conn$ itself, are given by
\begin{equation*}
\conn^\lp=\sum_{\vertex=0}^\infty \conn^{\lp,\vertex},\qquad
\conn=\sum_{\lp=0}^\infty \conn^\lp .
\end{equation*}
There is only one contribution with zero vertex number. This is the
Feynman propagator contributing to the 2-point function. Hence,
$\conn^{l,v}$ is zero if $v=0$ and $l\neq 0$, while $\conn^{0,0}$
is non-zero only on $V\tens V$ and coincides there with the Feynman
propagator. All non-zero vertex number contributions are captured by
the following corollary.

\begin{cor}
\label{cor:loopexpansion}
For $v\ge 1$:
\[
 \conn^{\lp,\vertex}=\fct^{\tens \vertex}\circ\Dopnew^{\lp,\vertex} .
\]
\end{cor}

\subsubsection*{Alternative recursion formula}
A key feature of $\Dopnew^{l,v}$ is that of satisfying an alternative  recursion relation. This has the advantage over (\ref{eq:recomega}),
that it may be translated directly into a recursion relation of the
resulting $n$-point functions $\conn^{l,v}$, related via
Corollary~\ref{cor:loopexpansion}.

\begin{prop}
\label{prop:alternativenew}
Let $v\ge 1$ and $l\ge 0$, but not $v=1$ and $l=0$. Then,
\begin{equation}
\Dopnew^{l,v}
=\frac{1}{l+v-1}\left(
\Dopnew^{l-1,v}\circ\Top
+ \sum_{a=0}^{l}\sum_{b=1}^{v-1}
\left(\Dopnew^{a,b}\tens\Dopnew^{l-a,v-b}\right)\circ\Qop \right)\,.\label{eq:altnew}
\end{equation}
The first summand does not contribute if $l=0$,
while the second does not contribute if $v=1$.
\end{prop}
Proposition \ref{prop:alternativenew} is proved by induction on  the number of internal edges $\edge=\lp+\vertex-1$  \cite{MeOe:loop}. Formula (\ref{eq:altnew}) has a
straightforward interpretation in terms of sums over weighted
graphs following the correspondence of Section~\ref{sec:algrep}.
Namely, the formula states that the weighted sum over graphs
with $l$ loops and  $v$ vertices is given by a sum of two
terms divided by $\edge$. 
The first term is the sum over all weighted graphs with
$l-1$ loops and  $v$ vertices which have an extra internal edge
attached, its end points being connected to vertices in all possible
ways.
The second term is a sum over
all ordered pairs of weighted graphs with total number of
vertices equal to $v$ and total number of loops equal to
$l$, connected in all possible ways with an internal edge.

Combining this result with Corollary~\ref{cor:loopexpansion} yields
the corresponding recursion formula for $\conn^{l,v}$.
\begin{cor}
\label{cor:recnew}
Let $v\ge 1$ and $l\ge 0$, but not $v=1$ and $l=0$. Then,
\begin{equation*}
\conn^{l,v}
=\frac{1}{l+v-1}\left(
\conn^{l-1,v}\circ\Top
+ \sum_{a=0}^{l}\sum_{b=1}^{v-1}
\left(\conn^{a,b}\tens\conn^{l-a,v-b}\right)\circ\Qop \right) .
\end{equation*}
It is understood that the first summand does not contribute if $l=0$,
while the second does not contribute if $v=1$.
\end{cor}

\section{Connected and 1PI $n$-point functions} \label{sec:conn1pi}
Now, we turn attention to simple algebraic expressions for the relation between connected and 1PI $n$-point
functions following from Theorem \ref{thm:main} and Corollary \ref{cor:recnew} \cite{MeOe:npoint}.

\begin{thm}
\label{thm:conn1pi}
The connected $n$-point functions $\conn$ may be expressed in terms of the 1PI
ones $\opi$ through the formula
\begin{equation*}
\conn=\conn^0+\sum_{\vertex=1}^\infty \conn^\vertex,\quad\text{with}\quad
 \conn^\vertex\defeq \opi^{\tens \vertex}\circ\Dopnew^{0,\vertex}\,,
%\label{eq:conn1pi}
\end{equation*}
where $\conn^0$ is non-zero only on $V\tens V$ and coincides there with the Feynman propagator.
\end{thm}
Since the 0-point and 1-point
1PI functions are zero, Theorem \ref{thm:conn1pi} holds as long as   all  tree graphs with $\vertex$ vertices, $n$ external edges whose end points are labeled by $x_1,\dots,x_n$, and the property that each vertex has
valence at least two, have no non-trivial symmetries. In other words, all tree graphs with the aforesaid properties are required to  occur in
$\Dopnew^{0,\vertex}(\phi(x_1)\cdots\phi(x_n))$  with exactly weight  $1$.  This  is ensured by the following lemma.

\begin{lem}
Consider a tree graph $\gamma$, all of whose vertices have valence at
least two. Then, $\gamma$ has no non-trivial symmetries.
\end{lem}
\begin{comment}
\begin{proof}
Consider a vertex $u$ of $\gamma$. We show that any symmetry must
leave $u$ invariant. If $u$ carries an external edge it must be
invariant since it is distinguishable. Thus, assume $u$ carries no
external edge. Choose one internal edge $e$ connected to $u$. Cut
$\gamma$ into
two by removing $e$. This yields two tree graphs $\gamma_1$ and
$\gamma_2$. Each of these must have at least one external edge to
satisfy the
valence requirement. Say $e_1$ is an external edge of $\gamma_1$ and
$e_2$ an external edge of $\gamma_2$. Since $\gamma$ is a tree there is
exactly one path to connect $e_1$ with $e_2$. Since the vertices
connected with $e_1$ and $e_2$ are held fixed under any symmetry so is
the whole chain of vertices formed by the path. However, $u$ is part
of this chain by construction and thus held fixed by any
symmetry.
\end{proof}
\end{comment}

In particular, by Corollary  \ref{cor:recnew}, the $\conn^\vertex$ satisfy the following recursion formula.

\begin{prop}
\label{prop:recconn}
$\conn^\vertex$ may be determined recursively via $\conn^1=\opi$ and with the
recursion equation for $\vertex\ge2$,
\[
\conn^{\vertex}=\frac{1}{\vertex-1}\sum_{i=1}^{\vertex-1}
 (\conn^i\tens \conn^{\vertex-i})\circ \Qop .
\]
\end{prop}

We notice that all results immediately carry over
to the relation between connected and modified 1PI $n$-functions. To this end, we modify the
definition of $\Rop$ (and consequently that of $\Dopnew^{0,\vertex}$) by replacing $\Gfeyn^{-1}$ with
$\Gconn^{(2)\,-1}$ in formula (\ref{eq:Rij}).
Recall that for the modified 1PI functions  $\Gopim$ not only $0$- and $1$-point functions vanish, but also $2$-point functions. This implies that only trees contribute which have
the property that all their vertices have valence at least
three. Actually, for a given number of external edges, there are only finitely many such trees. Therefore,  a connected  function yields a finite sum over tree graphs with modified 1PI functions as vertices,  for each  set of external edges, i.e., for each element of $\SV$ to which it is applied.

\subsection*{Acknowledgments}
We would like to thank the organizers of the Conference on Combinatorics and
Physics for their kind invitation. The research was supported in part by the
Czech Ministry of Education, Youth
and Sports within the project LC06002.

\bibliographystyle{amsplain}
\bibliography{bibliography}

\end{document}